\documentclass[longauth]{aa}
\usepackage{comment}
\usepackage[svgnames]{xcolor}
\usepackage{xstring}
\newcommand{\EatOneArg}[1]{}
\usepackage{siunitx}
\usepackage{url,hyperref}
\usepackage{amssymb,amsmath}
\usepackage{comment}
\usepackage[switch]{lineno}
\usepackage[justification=centering]{caption}

\hypersetup{                
    colorlinks=true,        
   linkcolor=blue,          
      citecolor=blue,        
      urlcolor=blue,           
}

\usepackage{graphicx}
\usepackage{dblfloatfix}
\usepackage{txfonts}
\begin{document}

   \title{Curvature in the very-high energy gamma-ray spectrum of M\,87}
   \date{\today}
   \author{H.E.S.S. Collaboration
    \and F.~Aharonian \inst{\ref{DIAS},\ref{MPIK},\ref{Yerevan_state}}
    \and F.~Ait~Benkhali \inst{\ref{LSW}}
    \and J.~Aschersleben \inst{\ref{Groningen}}
    \and H.~Ashkar \inst{\ref{LLR}}
    \and M.~Backes \inst{\ref{UNAM},\ref{NWU}}
    \and V.~Barbosa~Martins\protect\footnotemark[1] \inst{\ref{DESY}}
    \and R.~Batzofin \inst{\ref{UP}}
    \and Y.~Becherini \inst{\ref{APC},\ref{Linnaeus}}
    \and D.~Berge \inst{\ref{DESY},\ref{HUB}}
    \and K.~Bernl\"ohr \inst{\ref{MPIK}}
    \and M.~B\"ottcher \inst{\ref{NWU}}
    \and C.~Boisson \inst{\ref{LUTH}}
    \and J.~Bolmont \inst{\ref{LPNHE}}
    \and M.~de~Bony~de~Lavergne \inst{\ref{CEA}}
    \and F.~Bradascio \inst{\ref{CEA}}
    \and R.~Brose \inst{\ref{DIAS}}
    \and F.~Brun \inst{\ref{CEA}}
    \and B.~Bruno \inst{\ref{ECAP}}
    \and T.~Bulik \inst{\ref{UWarsaw}}
    \and C.~Burger-Scheidlin \inst{\ref{DIAS}}
    \and T.~Bylund \inst{\ref{CEA}}
    \and S.~Casanova \inst{\ref{IFJPAN}}
    \and R.~Cecil\protect\footnotemark[1] \inst{\ref{UHH}}
    \and J.~Celic \inst{\ref{ECAP}}
    \and M.~Cerruti \inst{\ref{APC}}
    \and T.~Chand \inst{\ref{NWU}}
    \and S.~Chandra \inst{\ref{NWU}}
    \and A.~Chen \inst{\ref{Wits}}
    \and J.~Chibueze \inst{\ref{NWU}}
    \and O.~Chibueze \inst{\ref{NWU}}
    \and T.~Collins \inst{\ref{UP}}
    \and G.~Cotter \inst{\ref{Oxford}}
    \and J.~Damascene~Mbarubucyeye \inst{\ref{DESY}}
    \and A.~Djannati-Ata\"i \inst{\ref{APC}}
    \and J.~Djuvsland \inst{\ref{MPIK}}
    \and A.~Dmytriiev \inst{\ref{NWU}}
    \and K.~Egberts \inst{\ref{UP}}
    \and S.~Einecke \inst{\ref{Adelaide}}
    \and S.~Fegan \inst{\ref{LLR}}
    \and G.~Fontaine \inst{\ref{LLR}}
    \and S.~Funk \inst{\ref{ECAP}}
    \and S.~Gabici \inst{\ref{APC}}
    \and J.F.~Glicenstein \inst{\ref{CEA}}
    \and J.~Glombitza \inst{\ref{ECAP}}
    \and G.~Grolleron \inst{\ref{LPNHE}}
    \and L.~Haerer \inst{\ref{MPIK}}
    \and W.~Hofmann \inst{\ref{MPIK}}
    \and T.~L.~Holch \inst{\ref{DESY}}
    \and M.~Holler \inst{\ref{Innsbruck}}
    \and D.~Horns \inst{\ref{UHH}}
    \and M.~Jamrozy \inst{\ref{UJK}}
    \and F.~Jankowsky \inst{\ref{LSW}}
    \and V.~Joshi \inst{\ref{ECAP}}
    \and I.~Jung-Richardt \inst{\ref{ECAP}}
    \and E.~Kasai \inst{\ref{UNAM}}
    \and K.~Katarzy{\'n}ski \inst{\ref{NCUT}}
    \and R.~Khatoon \inst{\ref{NWU}}
    \and B.~Kh\'elifi \inst{\ref{APC}}
    \and W.~Klu\'{z}niak \inst{\ref{NCAC}}
    \and Nu.~Komin \inst{\ref{Wits}}
    \and K.~Kosack \inst{\ref{CEA}}
    \and D.~Kostunin \inst{\ref{DESY}}
    \and A.~Kundu \inst{\ref{NWU}}
    \and R.G.~Lang \inst{\ref{ECAP}}
    \and S.~Le~Stum \inst{\ref{CPPM}}
    \and F.~Leitl \inst{\ref{ECAP}}
    \and A.~Lemi\`ere \inst{\ref{APC}}
    \and M.~Lemoine-Goumard \inst{\ref{CENBG}}
    \and J.-P.~Lenain \inst{\ref{LPNHE}}
    \and I.~Lypova\protect\footnotemark[1] \inst{\ref{LSW}}
    \and A.~Luashvili \inst{\ref{LUTH}}
    \and J.~Mackey \inst{\ref{DIAS}}
    \and D.~Malyshev \inst{\ref{IAAT}}
    \and G.~Mart\'i-Devesa \inst{\ref{Innsbruck}}
    \and R.~Marx \inst{\ref{LSW}}
    \and A.~Mehta \inst{\ref{DESY}}
    \and M.~Meyer\protect\footnotemark[1] \inst{\ref{UHH}}
    \and A.~Mitchell \inst{\ref{ECAP}}
    \and R.~Moderski \inst{\ref{NCAC}}
    \and M.O.~Moghadam \inst{\ref{UP}}
    \and L.~Mohrmann \inst{\ref{MPIK}}
    \and A.~Montanari \inst{\ref{LSW}}
    \and E.~Moulin \inst{\ref{CEA}}
    \and T.~Murach \inst{\ref{DESY}}
    \and M.~de~Naurois \inst{\ref{LLR}}
    \and J.~Niemiec \inst{\ref{IFJPAN}}
    \and S.~Ohm \inst{\ref{DESY}}
    \and L.~Olivera-Nieto \inst{\ref{MPIK}}
    \and E.~de~Ona~Wilhelmi \inst{\ref{DESY}}
    \and S.~Panny \inst{\ref{Innsbruck}}
    \and M.~Panter \inst{\ref{MPIK}}
    \and R.D.~Parsons \inst{\ref{HUB}}
    \and U.~Pensec \inst{\ref{LPNHE}}
    \and S.~Pita \inst{\ref{APC}}
    \and G.~P\"uhlhofer \inst{\ref{IAAT}}
    \and M.~Punch \inst{\ref{APC}}
    \and A.~Quirrenbach \inst{\ref{LSW}}
    \and M.~Regeard \inst{\ref{APC}}
    \and A.~Reimer \inst{\ref{Innsbruck}}
    \and O.~Reimer \inst{\ref{Innsbruck}}
    \and H.~Ren \inst{\ref{MPIK}}
    \and B.~Reville \inst{\ref{MPIK}}
    \and F.~Rieger \inst{\ref{MPIK}}
    \and B.~Rudak \inst{\ref{NCAC}}
    \and E.~Ruiz-Velasco \inst{\ref{MPIK}}
    \and V.~Sahakian \inst{\ref{Yerevan_institute}}
    \and H.~Salzmann \inst{\ref{IAAT}}
    \and A.~Santangelo \inst{\ref{IAAT}}
    \and M.~Sasaki \inst{\ref{ECAP}}
    \and F.~Sch\"ussler \inst{\ref{CEA}}
    \and H.M.~Schutte \inst{\ref{NWU}}
    \and J.N.S.~Shapopi \inst{\ref{UNAM}}
    \and H.~Sol \inst{\ref{LUTH}}
    \and S.~Spencer \inst{\ref{ECAP}}
    \and {\L.}~Stawarz \inst{\ref{UJK}}
    \and R.~Steenkamp \inst{\ref{UNAM}}
    \and S.~Steinmassl \inst{\ref{MPIK}}
    \and C.~Steppa \inst{\ref{UP}}
    \and K.~Streil \inst{\ref{ECAP}}
    \and T.~Takahashi \inst{\ref{KAVLI}}
    \and T.~Tanaka \inst{\ref{Konan}}
    \and A.M.~Taylor \inst{\ref{DESY}}
    \and R.~Terrier \inst{\ref{APC}}
    \and M.~Tsirou \inst{\ref{DESY}}
    \and C.~van~Eldik \inst{\ref{ECAP}}
    \and C.~Venter \inst{\ref{NWU}}
    \and J.~Vink \inst{\ref{Amsterdam}}
    \and T.~Wach \inst{\ref{ECAP}}
    \and S.J.~Wagner \inst{\ref{LSW}}
    \and A.~Wierzcholska \inst{\ref{IFJPAN}}
    \and M.~Zacharias \inst{\ref{LSW},\ref{NWU}}
    \and A.A.~Zdziarski \inst{\ref{NCAC}}
    \and A.~Zech \inst{\ref{LUTH}}
    \and P.~Zilberman\protect\footnotemark[1] \inst{\ref{DESY}, \text{now at }\ref{Perri}}
    \and N.~\.Zywucka \inst{\ref{NWU}}
    }

    \institute{
    Dublin Institute for Advanced Studies, 31 Fitzwilliam Place, Dublin 2, Ireland \label{DIAS} \and
    Max-Planck-Institut f\"ur Kernphysik, P.O. Box 103980, D 69029 Heidelberg, Germany \label{MPIK} \and
    Yerevan State University,  1 Alek Manukyan St, Yerevan 0025, Armenia \label{Yerevan_state} \and
    Yerevan Physics Institute, 2 Alikhanian Brothers St., 0036 Yerevan, Armenia
    \label{Yerevan_institute} \and
    Landessternwarte, Universit\"at Heidelberg, K\"onigstuhl, D 69117 Heidelberg, Germany \label{LSW} \and
    Kapteyn Astronomical Institute, University of Groningen, Landleven 12, 9747 AD Groningen, The Netherlands \label{Groningen} \and
    Laboratoire Leprince-Ringuet, École Polytechnique, CNRS, Institut Polytechnique de Paris, F-91128 Palaiseau, France \label{LLR} \and
    University of Namibia, Department of Physics, Private Bag 13301, Windhoek 10005, Namibia \label{UNAM} \and
    Centre for Space Research, North-West University, Potchefstroom 2520, South Africa \label{NWU} \and
    Deutsches Elektronen-Synchrotron DESY, Platanenallee 6, 15738 Zeuthen, Germany \label{DESY} \and
    Institut f\"ur Physik und Astronomie, Universit\"at Potsdam,  Karl-Liebknecht-Strasse 24/25, D 14476 Potsdam, Germany \label{UP} \and
    Université de Paris, CNRS, Astroparticule et Cosmologie, F-75013 Paris, France \label{APC} \and
    Department of Physics and Electrical Engineering, Linnaeus University,  351 95 V\"axj\"o, Sweden \label{Linnaeus} \and
    Institut f\"ur Physik, Humboldt-Universit\"at zu Berlin, Newtonstr. 15, D 12489 Berlin, Germany \label{HUB} \and
    Laboratoire Univers et Théories, Observatoire de Paris, Université PSL, CNRS, Université Paris Cité, 5 Pl. Jules Janssen, 92190 Meudon, France \label{LUTH} \and
    Sorbonne Universit\'e, Universit\'e Paris Diderot, Sorbonne Paris Cit\'e, CNRS/IN2P3, Laboratoire de Physique Nucl\'eaire et de Hautes Energies, LPNHE, 4 Place Jussieu, F-75252 Paris, France \label{LPNHE} \and
    IRFU, CEA, Universit\'e Paris-Saclay, F-91191 Gif-sur-Yvette, France \label{CEA} \and
    Friedrich-Alexander-Universit\"at Erlangen-N\"urnberg, Erlangen Centre for Astroparticle Physics, Nikolaus-Fiebiger-Str. 2, 91058 Erlangen, Germany \label{ECAP} \and
    Astronomical Observatory, The University of Warsaw, Al. Ujazdowskie 4, 00-478 Warsaw, Poland \label{UWarsaw} \and
    Instytut Fizyki J\c{a}drowej PAN, ul. Radzikowskiego 152, 31-342 Krak{\'o}w, Poland \label{IFJPAN} \and
    Universit\"at Hamburg, Institut f\"ur Experimentalphysik, Luruper Chaussee 149, D 22761 Hamburg, Germany \label{UHH} \and
    School of Physics, University of the Witwatersrand, 1 Jan Smuts Avenue, Braamfontein, Johannesburg, 2050 South Africa \label{Wits} \and
    University of Oxford, Department of Physics, Denys Wilkinson Building, Keble Road, Oxford OX1 3RH, UK \label{Oxford} \and
    School of Physical Sciences, University of Adelaide, Adelaide 5005, Australia \label{Adelaide} \and
    Universit\"at Innsbruck, Institut f\"ur Astro- und Teilchenphysik, Technikerstraße 25, 6020 Innsbruck, Austria \label{Innsbruck} \and
    Obserwatorium Astronomiczne, Uniwersytet Jagiello{\'n}ski, ul. Orla 171, 30-244 Krak{\'o}w, Poland \label{UJK} \and
    Institute of Astronomy, Faculty of Physics, Astronomy and Informatics, Nicolaus Copernicus University,  Grudziadzka 5, 87-100 Torun, Poland \label{NCUT} \and
    Nicolaus Copernicus Astronomical Center, Polish Academy of Sciences, ul. Bartycka 18, 00-716 Warsaw, Poland \label{NCAC} \and
    Aix Marseille Universit\'e, CNRS/IN2P3, CPPM, Marseille, France \label{CPPM} \and
    Universit\'e Bordeaux, CNRS, LP2I Bordeaux, UMR 5797, F-33170 Gradignan, France \label{CENBG} \and
    Institut f\"ur Astronomie und Astrophysik, Universit\"at T\"ubingen, Sand 1, D 72076 T\"ubingen, Germany \label{IAAT} \and
    Kavli Institute for the Physics and Mathematics of the Universe (WPI), The University of Tokyo Institutes for Advanced Study (UTIAS), The University of Tokyo, 5-1-5 Kashiwa-no-Ha, Kashiwa, Chiba, 277-8583, Japan \label{KAVLI} \and
    Department of Physics, Konan University, 8-9-1 Okamoto, Higashinada, Kobe, Hyogo 658-8501, Japan \label{Konan} \and
    GRAPPA, Anton Pannekoek Institute for Astronomy, University of Amsterdam,  Science Park 904, 1098 XH Amsterdam, The Netherlands \label{Amsterdam} \and
    Department of Physics, University of Wisconsin–Madison, Madison, WI 53706, USA \label{Perri}
    }
    
    \offprints{H.E.S.S.~collaboration,
    \protect\\\email{\href{mailto:contact.hess@hess-experiment.eu}{contact.hess@hess-experiment.eu}};
    \protect\\\protect\footnotemark[1] Corresponding authors
    }

 
   \abstract
   {
   The radio galaxy M\,87 is a variable very-high energy (VHE) gamma-ray source, exhibiting three major flares reported in 2005, 2008, and 2010. Despite extensive studies, the origin of the VHE gamma-ray emission is yet to be understood. In this study, we investigate the VHE gamma-ray spectrum of M\,87 during states of high gamma-ray activity, utilizing 20.2$\,$ hours the H.E.S.S. observations. Our findings indicate a preference for a curved spectrum, characterized by a log-parabola model with extra-galactic background light (EBL) model above 0.3$\,$TeV at the 4$\sigma$ level, compared to a power-law spectrum with EBL. We investigate the degeneracy between the absorption feature and the EBL normalization and derive upper limits on EBL models mainly sensitive in the wavelength range 12.4$\,$$\mu$m -- 40$\,$$\mu$m.
}

    \authorrunning{H.E.S.S. Collaboration et al.}

    \titlerunning{Curvature in the very-high-energy gamma-ray spectrum of M\,87}

    \maketitle
%

\section{Introduction}

Messier 87 (M\,87) is a bright nearby radio galaxy at 16.8$\,$Mpc from Earth \citep{Akiyama2019}, one of the rare radio galaxies from which TeV photons have been detected \citep{Aharonian2003}. This galaxy has been the subject of extensive study across the entire electromagnetic spectrum (we refer to \citet{EHT2017} for a review of recent observations). M\,87 is a variable VHE gamma-ray source with three major TeV flares reported to date; see \citet{2005flare}, \citet{2008flare}, and \citet{2010flare}. The potential origin of such rapid flares ($\sim$1 day) is linked to either the core of M$\,$87 or to very compact regions in the jet, for instance, the HST-1 knot \citep{Harris2006}.


Understanding the TeV radiation from radio galaxies helps shed light on the radiation mechanisms in blazars. Various mechanisms have been proposed to explain the production of very-high energy (VHE, $E$>100$\,$GeV) gamma-ray flares from M\,87. These include, e.g., leptonic models \citep[for a review, ][]{Rieger2018}, magnetospheric models \citep{Ansoldi2018}, multi-blob synchrotron self-Compton \citep[SSC; ][]{Lenain2008}, photo-hadronic model \citep{photohad_sah19, Alfaro2022}, a cloud-jet interaction model \citep{M87_flare_orig_Bar12}. Stochastic acceleration \citep{logpar_1,logpar_2} may lead, for instance, to the production of curved electron spectra, which ultimately lead to a curved gamma-ray spectrum. Close to the maximum particle energy in the accelerator, the resultant gamma-ray spectrum might follow a stretched cut-off shape, mimicking a curvature \citep{Romoli2017}. Furthermore, Klein-Nishina effects might also induce a cut-off in the gamma-ray spectrum at the highest energies \citep{Lefa2012}. Such features in the spectrum of the detected gamma rays might point to the mechanism and location of acceleration of the particles, as well as the type of relativistic particle. 

Regardless of the origin of the VHE flares from M\,87, the emitted gamma rays propagate to Earth, potentially interacting with photon fields en route, producing electron-positron pairs, and leaving an imprint in the spectrum. We identify three photon fields (and regions) that could significantly alter the VHE gamma-ray spectrum of M\,87. Firstly, internal absorption, i.e., the absorption of gamma rays at the core of M\,87 by the photons emitted in the accretion region \citep{M87-gamma-gamma-absorb}. Secondly, the gamma-ray absorption by the host galaxy's starlight \citep{Stawarz2006, Zacharias2017}, and thirdly, the extra-galactic background light (EBL) \citep[e.g., ][]{Franceschini2019}, all of which may leave detectable imprints in the observed spectrum of M\,87 \citep[e.g., Fig. 11.5 from ][]{Aharonian2004}. These imprints help us identify the photon fields in the neighborhood of the acceleration site, therefore, shedding light on the origin of the emission and on the mechanism of particle acceleration.

One approach to investigate spectral imprints is to examine the $\gamma$-$\gamma$ absorption of TeV photons by photon fields from the region. \citet{M87-gamma-gamma-absorb} suggest that the theoretical optical depth $\tau(E, z)$ from internal $\gamma$-$\gamma$ absorption reaches $\tau$$\gtrsim$1 at $\gtrsim$20$\,$TeV, assuming for that the VHE gamma rays are produced within 25$r_\mathrm{S}$ from the central black hole, where $r_\mathrm{S}$ is the Schwarzschild radius of the super-massive black hole (SMBH). This estimate assumes a Shakura-Sunyaev disk and a mass of the SMBH of $(6.4\pm0.5)\times10^9\, M_\odot$, where $M_\odot$ is the solar mass, which is in agreement with the latest estimates of the SMBH mass \citep{Akiyama2019}.

As we move further away from the SMBH, starlight becomes the dominant photon field. The galaxy extends up to a half-light radius of $R_\mathrm{1/2}$=7.2$\,$kpc \citep{Weil1997}, i.e., the radius within which half of the total starlight emitted by the galaxy is contained. By considering the presence of starlight from the host galaxy, one can estimate the effect of photon-photon absorption on the gamma-ray spectrum from the central AGN. \citep{Zacharias2017}. However, the lack of information about the spatial distribution of photons along the line of sight may complicate the interpretation of the results.

Beyond the M\,87 optical galaxy, the EBL becomes the dominant photon field in the $\mu$m wavelength range. The EBL encompasses all light emitted from stars and dust in galaxies throughout the Universe's history. It contains valuable information about stellar and galactic evolution. Accurate estimation of the EBL is crucial for studying extragalactic gamma-ray sources. Conversely, if an extragalactic source's intrinsic spectrum is well known, it can also provide a valuable avenue for estimating the EBL.

Attempts to estimate the EBL via attenuation of VHE photons primarily probe the 1$\,\mu$m -- 50$\,\mu$m range (see e.g. \citet{Aharonian2007, HESS_EBL13, MAGIC2016, HESS_EBL, MAGIC2019, Abeysekara2019} or \citet{review_galaxies} for a review). This range overlaps with the range probed by direct measurements. Since direct measurements suffer from foreground contamination \citep{Hauser1998}, VHE gamma-ray absorption by the EBL provides complementary, indirect estimates. 

The VHE flux from gamma-ray sources quickly falls off at higher energies, decreasing typically as a power law (PL) with photon index $\Gamma$$\sim$2 -- 3. Given that the optical depth $\tau$ increases monotonically with the gamma-ray energy, efforts to investigate its influence on the gamma-ray spectrum at the highest energies are restricted to bright and nearby VHE gamma-ray sources.

Due to its close proximity to Earth, M\,87 is a unique object to probe the 10$~\mu\mathrm{m}$ to 50$~\mu\mathrm{m}$ region of the EBL spectrum. According to several models, the theoretical optical depth should become measurable in the spectrum of M\,87 at $\gtrsim$ 10$\,$TeV \citep[e.g., Fig. 11.5 from ][]{Aharonian2004, Franceschini2019}. There is an observable hardening of the spectrum of the source during high states \citep[$\Gamma$$\approx$2.2, ][]{2010flare, Victor} which presents us with an opportunity to probe the EBL even at the source's relative proximity.


The High Energy Stereoscopic System (H.E.S.S.) has good coverage of all the aforementioned TeV gamma-ray flares, which makes it the ideal instrument to study the VHE gamma-ray spectrum of M\,87 up to the highest energies. Therefore, we select high-state data sets from H.E.S.S. observations of M\,87 and search for absorption imprints in it. The VHE gamma-ray observations of M\,87 can also be used to search for physics beyond the Standard Model. Using the same data set, \citet{Rahul} searched for spectral signatures of oscillations between photons and axion-like particles in the magnetic field of the Virgo Cluster. No significant evidence for such oscillations was found.

This publication is organized as follows: in Sect. \ref{sec:hess} we introduce the H.E.S.S. experiment, data set, and the methods used in the data analysis. In Sect. \ref{sec:results}, we present the results of the spectral fits, and in Sect.~\ref{sec:discussions}, we discuss the implications of our findings, before summarizing them in Sect.~\ref{sec:summary}.


\section{Observations and Data Analysis}
\label{sec:hess}

H.E.S.S. is an array of five Imaging Atmospheric Cherenkov Telescopes (IACTs) situated in the Khomas Highland of Namibia. The H.E.S.S. telescopes have been regularly used, initially with four telescopes, to observe M\,87 since 2004, both in monitoring campaigns and in response to triggers from external instruments reporting flaring activities. Observations are conducted in individual runs, each typically lasting up to 28 minutes. This study utilizes H.E.S.S. observations of M\,87 from 2004 to 2020. Although some of the data were already used in previous publications \citep{2005flare, 2008flare, 2010flare, EHT2017, Victor}, we focus here, for the first time, specifically on the high state, combining data from individual observational runs with elevated flux. To have a consistent data set, we only consider data from the 4 telescopes with 12$\,$m diameter mirrors (CT$\,$1--4), which have been in operation since the earliest detected flare. Given that the 28$\,$m diameter mirror telescope (CT$\,$5) is in operation only since 2012, it has not observed the aforementioned flaring states of M\,87. Hence, we choose to exclude its data from this study, ensuring a consistent representation of M\,87 in VHE gamma rays over time. We have applied the quality criteria for a spectral analysis \citep{crabhess2006}, selecting only the best quality observations with at least three telescopes participating in the observation.

The data were processed using \textit{gammapy v.1.1} \citep{gammapy2023}, based on the generation of FITS files \citep{Nigro2019} according to a multivariate gamma-selection and geometrical event reconstruction \citep{crabhess2006, boost_tree}. We select gamma-ray-like events within a cone of radius $\leq0.11^{\circ}$ around the position of the core of M87 at right ascension $187.7059^\circ$ and declination $12.3911^\circ$ for J2000 \citep{M87position}. The background is estimated utilizing the reflected-region background method \citep{background_techniques} with a minimum angular separation to M\,87 of 0.15$^\circ$. 

\begin{figure}
\centering
\includegraphics[width=\textwidth/2]{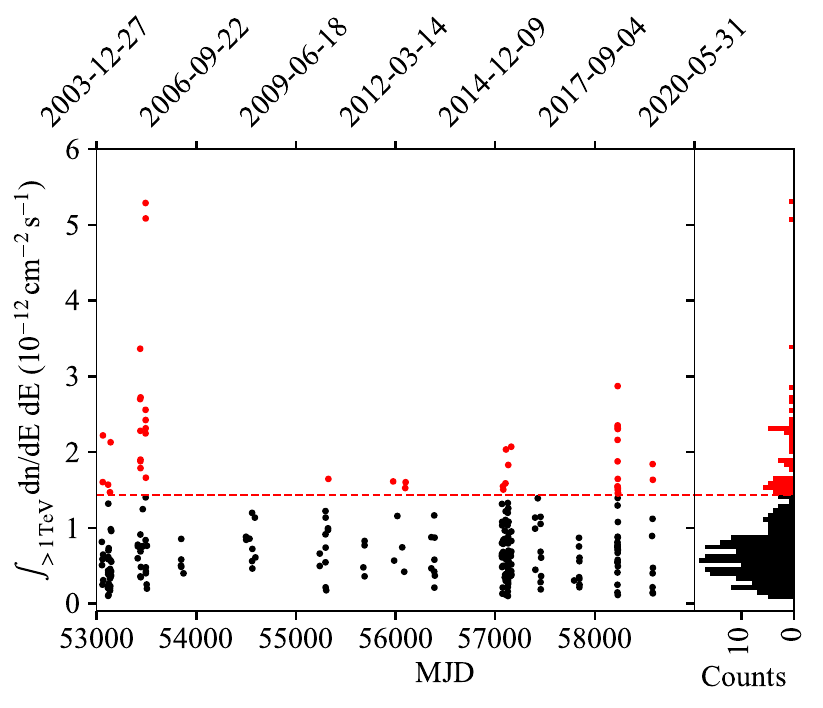}
\caption{Gamma-ray flux above 1\,TeV per observation run. The high state is defined by the 10\% highest flux above 1\,TeV, as indicated by the red data points (above the 90\% percentile of the distribution). In the inset on the right side, a histogram with number of observational runs below and above the 90\% percentile is shown.}
\label{fig:event_rate}
\end{figure}

In Fig.~\ref{fig:event_rate}, we present the run-wise flux of the selected data above 1$\,$TeV, emphasizing observational runs with elevated flux. All data points exhibiting a flux exceeding $10^{-13}$ cm$^{-2}$ s$^{-1}$ are included. For clarity, uncertainties associated with the flux points are omitted in order to maintain a clean visualization. It is important to note that each flux point does not necessarily denote a detection but rather represents the estimated flux for each observation. Based on an empirical cumulative distribution function, we select 10\% of the runs with the highest flux, i.e., the runs with the flux above the 90\% percentile of the distribution to compose the high state. The choice of the 90\% percentile is based on an expected effective duty cycle of the source, which is also influenced by the H.E.S.S. observation strategies in the past. The high state runs are marked in red in Fig.~\ref{fig:event_rate}. This amounts to 20.2 hours of collected data. The definition of high state ensures that we selected only the observations from the tail of the high-state flux distribution (see Fig~\ref{fig:event_rate} inset on the right panel). 

We regard the selection of the high state based on the 10\% highest flux above 1$\,$TeV as stringent, evidenced by the exclusion of the 2008 flare from our selection, which H.E.S.S. did not observe at its peak intensity. Some observation runs conducted during the 2010 flare were excluded from the dataset. This exclusion was due to these observations being performed on the target without the usual wobbling motion. The typical wobble is essential for obtaining reliable background and spectral estimates \citep{background_techniques}. In 2018, M\,87 underwent a VHE gamma-ray flare, which is to be reported in depth in a forthcoming publication within the Event Horizon Telescope Multiwavelength Group. Hence, our high state dataset predominantly comprises data from the 2005 and the 2018 flares. As a systematic check, we study the influence of the high state selection into the final spectral results and the rise of the curvature in the high state data set as shown in Appendix~\ref{app:spectra}.


We use the likelihood $\mathcal{L}$=$\prod_{i}P(n_i|\mu_i)$, and $P(n_i|\mu_i)$ as the Poisson probability of observing $n_i$ events in energy bin $i$, given the predicted number of events $\mu_i$ from the background and the source models, to fit the data to spectral models. We explore several spectral models to describe the differential gamma-ray flux $\phi(E)$ as a function of energy $E$, each with increasing complexity. The first model is a power law (PL) spectrum, defined as:

\begin{equation}
\label{eq:PL}
\phi(E) = \phi_0 \left( \frac{E}{E_0} \right)^{- \Gamma},
\end{equation}

\noindent where the normalization $\phi_0$ and the spectral index $\Gamma$ are free parameters, while the reference energy $E_0$ is fixed to the decorrelation energy, i.e., to the energy at which the correlation between the spectral model parameters is minimized.

To account for potential curvature in the spectrum, we consider a Log-Parabola (LP) spectral model:

\begin{equation}
\phi(E) = \phi_0 \left( \frac{E}{E_0} \right)^{- \Gamma - \beta \mathrm{ln}(E/E_0)},
\end{equation}

\noindent where the coefficient $\beta$, the normalization $\phi_0$, and the spectral index $\Gamma$ are free parameters.

To incorporate the EBL absorption into our spectral models, we initially reviewed a range of models from the literature, eventually selecting three that comprehensively represent the range of options. We started with the \textit{kneiske} model \citep{Kneiske2010}, tailored to emulate the lower boundary of the EBL flux. For a contemporary estimate of the EBL founded on data, we turned to the \textit{finke2022} model, specifically referencing model A from \citet{Finke2022}. To round off our selection with a model reflecting high-intensity EBL, we incorporated the upper bounds of the Dominguez model, termed \textit{dominguez-upper}, which mirrors the upper uncertainties outlined in \citet{Dominguez2011}. Collectively, these three models capture the diverse models from literature regarding the optical depth $\tau$.

The models characterizing the EBL spectrum, presented as the optical depth $\tau$ in relation to the gamma-ray energy, are commonly expressed as functions of redshift. We used the tables from \citet{manuel} considering the redshift of M\,87 as $z$$\approx$0.0042. However, considering the potential influence of the local Hubble flow on the redshift measurements of nearby sources, we also took into account a direct distance measurement to M87 of 16.8$\,$Mpc \citep{Bird2010, Akiyama2019}. This distance was converted to a redshift z=0.00389 using the \textit{Planck18} cosmology \citep{Planck} with the assistance of \textit{Astropy v.5.3.3} \citep{astropy:2013}. The resulting variations between the two redshift values in terms of EBL optical depth are below 10\%. We base our estimates on the originally estimated redshift.

Finally, we incorporate an EBL component into the PL and LP spectral models and define the notation of the new models as PLxEBL and LPxEBL spectral models, respectively:

\begin{equation}
    \phi(E) = \phi_0 \left( \frac{E}{E_0} \right)^{- \Gamma} \exp(- \alpha_{\text{norm}}\tau(E,z)),
\end{equation}

\begin{equation}
    \phi(E) = \phi_0 \left( \frac{E}{E_0} \right)^{- \Gamma - \beta \mathrm{ln}(E/E_0)} \exp(- \alpha_{\text{norm}}\tau(E,z)),
\end{equation}

\noindent where $\tau(E, z)$ is the model-dependent EBL optical depth, parameterized by the photon energy and redshift, and $\alpha_{\text{norm}}$ is the EBL normalization. While $\tau(E, z)$ is defined by the EBL model, $\alpha_{\text{norm}}$ can either be set to vary freely, in case one is interested in constraining the EBL, or be fixed to the expected value $\alpha_{\text{norm}}$=1 to probe the intrinsic spectral model. As we explain in Sect.~\ref{sec:results}, we utilize both approaches. 

We used the implementation from \textit{gammapy} to fit the spectral models to the data following the forward-folded approach \citep{Piron2001} and obtain the best-fit parameters with their loglikelihood values ($-2\ln$\(\mathcal{L}\)). For nested models, the preference of a model~1 in comparison to model~0, where model~1 has more free parameters, is given by the test statistic $TS=-2\ln(\mathcal{L}_0/\mathcal{L}_1)$. For large event statistics, $TS$ follows a $\chi^2$ distribution with $k$ degrees of freedom, where $k$ is the difference of free parameters between the two nested models. This representation allows us to express the results in terms of the significance of the fit. Since $k$=$1$ for the PL and the LP spectral model comparison, the significance of the LP in comparison to the PL spectral model is approximated by $\sqrt{TS}$.

In addition to the main analysis described in this section, we performed two cross-checks to assess the reliability of the results. The first cross-check utilized an alternative high-state definition, based on \citet{Victor}. This approach led to a slightly different selection of observations, though the resulting best-fit spectral models are within 1$\sigma$ statistical uncertainties consistent with each other. The second cross-check utilized the same observations as the first cross-check with an independent analysis chain \citep{M++}. The results were also found to be consistent with the main analysis.

\section{H.E.S.S. analysis results}
\label{sec:results}


For the high-state dataset defined in Sect.~\ref{sec:hess}, we obtained 390$\pm$28 excess gamma-ray events and 376$\pm$5 expected background events for a livetime of 20.2 hours. The significance of detection is calculated based on Eq.~17 of \citet{Lima} and yields 16.9$\,$$\sigma$. For energies above 10$\,$TeV, we find 15$\pm$4 gamma-ray excess events for 2.3$\pm$0.4 expected background events, which leads to a significant detection of 6.0$\,\sigma$.

\begin{figure}
  \centering
  \resizebox{0.49\textwidth}{!}{\includegraphics{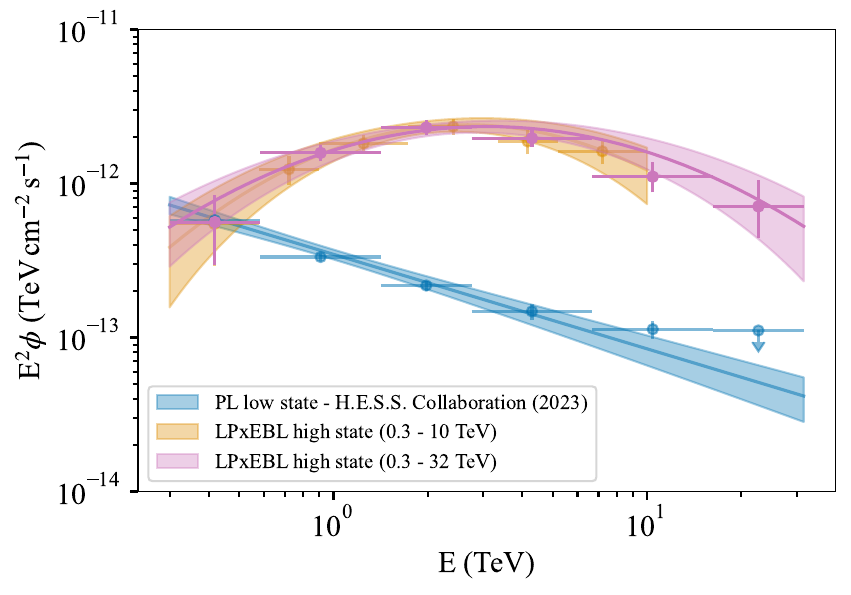}}
  \caption{Detected spectral energy distribution of M\,87 data with H.E.S.S. The best-fit LPxEBL \textit{finke2022} spectral models are shown in yellow for the reduced energy range ($0.3 - 10\,$TeV) and in purple for the full energy range ($0.3 - 32\,$TeV). The best-fit PL spectrum for the low-state data \citep[based on ][]{Victor} is shown for comparison in blue. The color bands show the 1$\,\sigma$ uncertainty contours. The upper limit (UL) in the last bin of the low-state spectrum is defined at a 95\% confidence level (c.l.).}
  \label{fig:spectrum} 
\end{figure}

Given the distance to M\,87, we find that the optical depth $\tau$<$\,$0.2 holds for energies <$\,$10$\,$TeV for all three considered EBL models. We consider the effects of EBL absorption for $\tau$<$\,$0.2 as small and focus first on the analysis of the gamma rays with energy between 0.3$\,$TeV and 10$\,$TeV. Based on the methodology described in Sect.~\ref{sec:hess}, we fit the PLxEBL \textit{finke2022} and the LPxEBL spectral models to the data in the reduced energy range. The results, shown in the first two rows of Table~\ref{table:all_fit_results} and in yellow in Fig.~\ref{fig:spectrum}, indicate a preference for an LPxEBL model over the PLxEBL model with a significance of 3.5$\,$$\sigma$. We further discuss the implications of this curvature in the following section.

In a second step, we consider the full energy range (0.3$\,$TeV$\,$--$\,$32$\,$TeV). We fit PLxEBL and LPxEBL models now for the three EBL models \textit{kneiske}, \textit{finke2022}, and \textit{dominguez-upper}, with fixed $\alpha_\mathrm{norm}$=1 to the data. The results are given in Table~\ref{table:all_fit_results} for all the models and in Fig.~\ref{fig:spectrum} by the red curve and data points. The curvature in the spectrum for the full energy range (0.3 - 32$\,$TeV) is detected, given that the LPxEBL models are preferred over the PLxEBL models with 4.4$\,$$\sigma$, 4.2$\,$$\sigma$, and 3.6$\,$$\sigma$ for the \textit{kneiske}, \textit{finke2022}, and \textit{dominguez-upper} EBL models, respectively. For the sake of readability, in the following discussions, we primarily refer to the \textit{finke2022} LPxEBL model, as it stands as the most recent and representative EBL model among the three considered. The energy-integrated flux above 0.3$\,$~TeV amounts to $(5.1\pm0.5)\times10^{-12}$cm$^{-2}$s$^{-1}$ for the \textit{finke2022} LPxEBL model.

\begin{table*}[]
    \centering
    \begin{tabular}{c c c c c c c }
        \hline
        \hline
            Model &
            -2ln\(\mathcal{L}\) &
            $\mathrm{\phi_0}$ &
            \textbf{$\Gamma$} &
            $\beta$ & 
            Ref. energy & 
            Energy range
            \\
            \hline
            PLxEBL \textit{finke2022} & 30.68 & 4.7$\pm$0.3 & 1.81$\pm$0.08 & - & 1.96 & 0.3 -- 10 \\
            LPxEBL \textit{finke2022} & 18.69 & 6.0$\pm$0.5 &  1.76$\pm$0.10 & 0.38$\pm$0.13 & 1.96 & 0.3 -- 10 \\
            \hline \hline
            PLxEBL \textit{kneiske} & 36.23 & 3.7$\pm$0.2 & 1.94$\pm$0.06 & - & 2.15 & 0.3 -- 32\\
            LPxEBL \textit{kneiske} & 16.86 &  4.8$\pm$0.4 & 1.81$\pm$0.08 & 0.29$\pm$0.07 & 2.15 & 0.3 -- 32 \\
            \hline \hline
            PLxEBL \textit{finke2022} & 34.38 & 3.8$\pm$0.2 & 1.92$\pm$0.06 & - & 2.15 & 0.3 -- 32\\
            LPxEBL \textit{finke2022} & 17.04 & 4.9$\pm$0.4 & 1.80$\pm$0.08 & 0.27$\pm$0.08 & 2.15 & 0.3 -- 32 \\
            \hline \hline
            PLxEBL \textit{dominguez-upper} & 31.28 & 3.9$\pm$0.2 & 1.88$\pm$0.06 & - & 2.15 & 0.3 -- 32\\
            LPxEBL \textit{dominguez-upper} & 18.20 & 4.9$\pm$0.4 & 1.79$\pm$0.08 & 0.25$\pm$0.08 & 2.15 & 0.3 -- 32\\
            \hline
            
    \end{tabular}
    \caption{Best fit results for the spectral VHE gamma-ray distribution of the high state of M\,87. We compare the PL and the respective LP best-fit spectral model for the reduced energy range and for the full energy range for each EBL model considered. The energy range is given in TeV and $\mathrm{\phi_0}$ is given in units of 10$^{-13}$cm$^{-2}$ s$^{-1}$TeV$^{-1}$. The decorrelation energy of the PL spectral model was used as the reference energy, given in TeV in the table.}
    \label{table:all_fit_results}
\end{table*}

Lastly, we evaluated the LPxEBL spectral model, as previously defined, in conjunction with the same three EBL models and with the EBL normalization $\alpha_\mathrm{norm}$ allowed to vary freely. Upon fitting the LPxEBL with a variable $\alpha_\mathrm{norm}$ to the full energy range dataset, the fit converged to a pure LP model. This suggests a degeneracy between the curvature parameter $\beta$ and the EBL normalization $\alpha_\mathrm{norm}$.

\begin{figure}
  \centering
  \resizebox{0.49\textwidth}{!}{\includegraphics{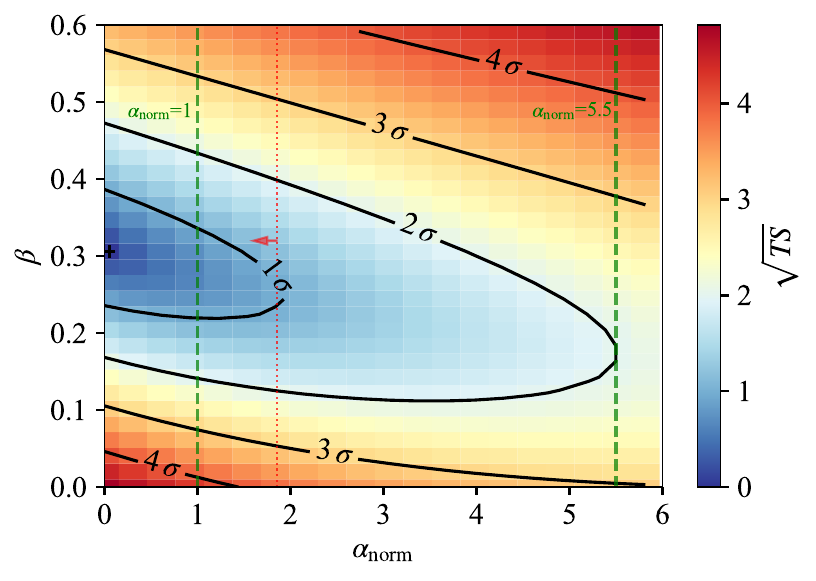}}
  \caption{Parameter space around the best-fit LPxEBL spectral model for the $\alpha_\mathrm{norm}$ and $\beta$ parameters of the \textit{finke2022} EBL model. The color map gives the distance in standard deviations from the best-fit parameters (shown by the black cross), i.e., the lowest value provides the best fit. The black contours give the 1$\,$$\sigma$, 2$\,$$\sigma$, 3$\,$$\sigma$, and 4$\,$$\sigma$ uncertainty contours. The vertical green lines show the $\alpha_\mathrm{norm}$=1 and $\alpha_\mathrm{norm}$=5.5, which provides the 95\% c.l. UL on the $\alpha$ parameter. For comparison, the dotted red line shows the $\alpha_\mathrm{norm}$ that corresponds to the UL above which the \textit{finke2022} EBL model would start to contradict the ULs from \citet{Biteau2015}.}
  \label{fig:contour}
\end{figure}

To elucidate the observed degeneracy, we present the log-likelihood values for the parameter space of $\alpha_\mathrm{norm}$ and $\beta$ within the LPxEBL model. The result for the \textit{finke2022} EBL model is shown in terms of $\sqrt{TS}$ from the best fit position in Fig.~\ref{fig:contour}. Based on the intersection of the $\alpha_\mathrm{norm}$ with the 2$\,\sigma$ contour in Fig.~\ref{fig:contour}, we derive the maximum allowed EBL normalization $\alpha_\mathrm{norm}$ within 95\% c.l. of $\alpha_\mathrm{norm}$<5.5, for the \textit{finke2022} EBL model. Similarly, for the \textit{kneiske}, and \textit{dominguez-upper} EBL models, we derive $\alpha_\mathrm{norm}$<8.7 and $\alpha_\mathrm{norm}$<2.0, respectively. In the case $\beta$=0, the best-fit position in the parameter space is always more than 4$\,\sigma$ away from the fit for an EBL intensity for $\alpha_\mathrm{norm}$$\lesssim$1.5. Since the gamma-ray energy range used is between 0.3$\,$TeV and 32$\,$TeV, our limits on the EBL intensity probe the $\approx0.4\,\mu m-\,40\mu$m EBL photons\footnote{This wavelength range is obtained from the peak of the pair production cross section at the wavelength $\lambda_\mathrm{max.}$$\approx$1.24($E_\mathrm{\gamma}$/TeV) $\mu$m, where $E_\mathrm{\gamma}$ is the gamma-ray energy \citep{Franceschini2019}.}, although our analysis is more sensitive for gamma rays $\gtrsim$10$\,$TeV, i.e., for EBL photons $\gtrsim$12.4$\mu$m, where their optical depth reaches $\tau$$\gtrsim$0.2.

The curvature observed in the spectrum of the high state of M\,87 was analyzed in this study using a LP distribution. Nevertheless, it is worth noting that an exponential cut-off model, despite having an equivalent number of degrees of freedom, may suggest different underlying physics. Hence, in Appendix~\ref{app:EC}, we explore the PL with an exponential cut-off, defined as the PLxECxEBL model, taking into consideration the effects of the EBL.

\section{Discussion and conclusions}
\label{sec:discussions}


A curved gamma-ray spectrum has been already observed up to TeV energies in the radio galaxy NGC1275 \citep{Ansoldi2018}, as well as in other gamma-ray sources \citep{Zdziarski2020}. Various mechanisms could underlie this spectral feature, suggesting its potential common occurrence in nature. In this section, we analyze possible physics scenarios that could lead to a curvature in the VHE gamma-ray spectrum of the high state of M$\,$87. We consider four possibilities and discuss them in the subsequent paragraphs based on the details of our analysis:

\begin{itemize}
  \item External absorption: EBL photons;
  \item Multi-component high state: a composition of different particle populations;
  \item Internal absorption: by local photon fields, e.g., the accretion flow and the star light;
  \item Intrinsic curved electron spectrum: due to the nature of the mechanism of acceleration, e.g., stochastic acceleration, or due to a cut off in the electron spectrum.
\end{itemize}

The presence of the curvature in the high-state spectrum of M\,87 at VHE energies in the reduced energy range ($0.3 - 10\,$TeV; first two rows in Table~\ref{table:all_fit_results}) indicates that the curvature is not related to the EBL absorption, which is negligible in this energy range. This hypothesis gained further support from the minimal discrepancies observed when fitting the spectral data within the reduced energy range, taking into account the EBL absorption within the model. Although showing a compatible curvature, the spectral curvature is more strongly detected in the full energy range ($0.3 - 32\,$TeV; Table~\ref{table:all_fit_results}). Hypotheses with more complex spectral behavior cannot be tested due to the limited event statistics in the dataset (Appendix~\ref{app:EC}).

For the full energy range ($0.3 - 32\,$TeV), we find that the EBL normalization ($\alpha_{\text{norm}}$) and the intrinsic curvature coefficient from the LP model ($\beta$) are statistically degenerate. However, given that the EBL is in fact directly obtained from observations at optical and infrared wavelengths, \citep[e.g., ][]{Dominguez2011, new_EBL}, the EBL normalization $\alpha_{\text{norm}}$ is expected to be at least close to one and, therefore, it cannot be set to zero in expense of $\beta$ as Fig.~\ref{fig:contour} might suggest. It is clear through the 4$\,$$\sigma$ statistical uncertainty contours from Fig.~\ref{fig:contour} that the curvature in the spectrum is strongly detected at a 4$\,$$\sigma$ level even by considering $\alpha_{\text{norm}}$=1. Therefore, based on current available EBL models, we rule out the external absorption by EBL photons as the exclusive cause for the observed curvature in this study. The existing EBL models utilize data collected from all directions in the sky and do not consider potential anisotropies in the EBL at the same redshift. This lack of consideration may influence the predicted absorption feature in gamma-ray spectra. \citet{voids}, for instance, examine but do not find a strong correlation between the location of voids and hard gamma-ray sources. However, they estimate a 2\% variation in the density of EBL photons from voids compared to the nominal EBL density. Notably, EBL anisotropy remains a relatively unexplored subject \citep{Hervet:2023wtn}.

Since our data set is a collection of high emission states, the overall VHE gamma-ray emission from the high state of M\,87, as defined here, results from the contribution of the individual flares. Although the flaring states of M\,87 are usually accompanied by a hard spectral index ($\Gamma$$ \approx$2), it is not trivial to conclude from it that the individual flares originate from the same region \citep{2010flare}. Therefore, different emission regions with distinguished electron populations might be behind the signal. We tested this hypothesis by fitting the data around the 2005 flare and the 2018 flare, as they are the largest contributors to the selected high state data (Fig.~\ref{fig:event_rate}), separately. The results show a hint of 1.5$\,\sigma$ and 2.9$\,\sigma$ preference for an LPxEBL \textit{finke2022} model in comparison to the PLxEBL \textit{finke2022} model, respectively, in the 2005 and 2018 flares. Furthermore, the spectral indices of the PL models from the 2005 and the 2018 flares are 2.01$\pm$0.09 and 1.9$\pm$0.1, respectively, which is consistent with each other. Therefore, the hint of curvature, especially from the 2018 dataset, and the similar spectral shapes suggest that the curvature detected in the combined dataset is not due exclusively to the superposition of different spectral components.

The presence of a persistent steady emission component even during flaring episodes could also contribute to the spectral shape, especially in its lower energy part ($E$$\approx$0.3$\,$TeV). In fact, the low state of M\,87 is well-described by a PL distribution with a steep spectral index \citep[$\Gamma$=2.63$\pm$0.09]{Victor}, where no significant spectral curvature was found in the H.E.S.S. data. If the low and high states of the VHE gamma-ray emission of M\,87 are indeed due to different components, the high-state component should be curved and dominate over the steady PL component during flares to yield the observed SED. However, the low-state component would still contribute, at the same level as the high-state component to the flux of gamma rays with energies around $E$$\approx$0.3$\,$TeV during flaring activities (see Fig.~\ref{fig:spectrum}). This hypothesis is also supported by the fact that \textit{Fermi}-LAT observations did not detect a significant increase in the high-energy (100$\,$MeV$\,$<$\,$$E$$\,$<$\,$300$\,$GeV) flux during the 2010 VHE gamma-ray flare \citep{2010flare}. Furthermore, the \textit{Fermi}-LAT flux points (regular state in \citet{fermi_states}) smoothly connect to the H.E.S.S. flux points of the low state PL distribution at $\approx$0.3$\,$TeV. Therefore, even in high emission states, we expect an up-rise in the curvature (Fig.~\ref{fig:app:spectra}) for energies below 0.3$\,$TeV, where the low-state would dominate the high-state component. The previous arguments support the hypothesis that the VHE gamma-ray emission of M\,87 is composed separately of a low-energy and a high-energy component, whereas the VHE high-energy component appears during flaring activities. Nonetheless, a single emission region responsible for the emission in both source emission states cannot yet be discarded.

As for the internal absorption, \citet{M87-gamma-gamma-absorb} estimates that the inner region of M\,87 should be transparent to gamma rays with energies $\lesssim$20$\,$TeV. The absorption by the starlight from the host galaxy can also be estimated based on its emissivity. We employ the surface brightness of M\,87 (UGC 7654) in the 1.4$\,$$\mu m$ -- 1.7$\,$$\mu$m range, denoted as $\mu_\lambda$=15.51 mag/arcsec$^2$ from \citet{Capetti2006}. Using Eq.~A5 of \citet{Zacharias2017}, we convert the surface brightness to emissivity $j_\lambda$:

\begin{equation}
    j_\lambda = 2.32\times 10^{-5}\left ( \dfrac{r_0}{\mathrm{cm}}\right )^{-1}10^{0.4\times(25.992 - (\mu_\lambda - A))} \left (\dfrac{\mathrm{erg}}{\mathrm{cm^{3}s^{1}}} \right ),
\end{equation}

\noindent where we set the galactic extinction $A$ to zero, as it has already been corrected for the $\mu_\lambda$ estimate and $r_0$ is the characteristic radius of the source. We use $r_0\approx 432\,$pc, half the size of the image in \citet{Capetti2006} used for the brightness estimate, converted to parsec by considering the distance to M\,87 of 16.8$\,$Mpc. For a radial distribution of light, the optical depth $\tau_{M87,\lambda}$ for the galactic emission up to $r_0$ centered at $\lambda$ is estimated through:

\begin{equation}
    \tau_{M87,\lambda} = \dfrac{\sigma_{\gamma \gamma} j_\lambda \lambda r_0^2}{h c^2},
\end{equation}

\noindent where $\sigma_{\gamma\gamma}$ is the cross section for $\gamma-\gamma$ absorption, $c$ is the speed of light, and $h$ is the Planck constant. We utilized the full cross section \citep{PhysRev.155.1404} for an average cosinus angle of interaction between the incoming gamma rays and the photons from the star light of 0.5. Our estimates indicate that the galactic optical depth at $\lambda$=1.6$\,\mu$m peaks for a gamma-ray energy of $\approx$1$\,$TeV, although $\tau_{M87,\lambda}$<0.001 in the entire gamma-ray energy range, i.e. from the threshold energy of pair production up to 32\,TeV. Therefore, gamma-ray absorption by starlight is weak and unlikely to explain the reported curvature here.


The curvature in the gamma-ray spectrum can also be associated with an intrinsically curved particle spectrum, typically explained by stochastic particle accelerations \citep{logpar_1,logpar_2}. The event statistics of our dataset do not allow us to distinguish between a cutoff on the electron spectrum due to the maximum electron energy in the source or a cutoff due to the onset of the Klein-Nishina regime \citep{kleinnishina} from an intrinsic parabolic curvature. The rapid VHE gamma-ray flares undergone by M\,87 with hard spectral indices point to an efficient acceleration mechanism. However, we are not able to distinguish the acceleration mechanism based solely on the spectral shape of the VHE gamma rays. In addition to that, the exact acceleration site also remains an open question \citep[see ][]{Rieger2012}.

\section{Summary}
\label{sec:summary}
In this work, we study the spectrum of the high-state emission of M\,87 with H.E.S.S. data. We define the high state as the 10\% of the observation runs with the highest flux above 1$\,$TeV (Fig.~\ref{fig:event_rate}). This accounts for a total of 20.2$\,$ hours of observations.

We first consider the energy range between 0.3 and 10$\,$TeV, in which the EBL absorption is small, i.e., $\tau$<0.2. We then fit a power-law (PL) distribution with the \textit{finke2022} \citep{Finke2022} EBL model, as the null hypothesis, to describe the spectrum and test alternative spectral models with curvature. The result shows a strong (4.2$\,\sigma$) preference for a log parabola (LP) spectral model with the EBL (LPxEBL) in comparison to the PL model with EBL (PLxEBL), which establishes a curved spectrum in the VHE gamma-ray emission from the high state of M\,87. 

In a second analysis step, we consider the full energy range ($0.3$ -- $32\,$TeV) and fit an LPxEBL to the data. We utilized three EBL models at $z\approx0.0042$ that cover a multitude of other models. Therefore, our selection is representative \citep{manuel}: the \textit{kneiske} model \citep{Kneiske2010}, the \textit{finke2022} model \citep[model A in ][]{finke_justin_d_2022_7023073}, and the \textit{dominguez-upper} model \citep[the upper uncertainties in ][]{Dominguez2011}. We confirm that the curvature in the spectrum is also strongly detected considering the full energy range as the LPxEBL models are preferred at $\approx$4$\,$$\sigma$ over the PLxEBL models. The total flux resulted from the analysis above 0.3$\,$TeV is estimated to be (5.1$\pm$0.5)x10$^{-12}$cm$^{-2}$s$^{-1}$ for the \textit{finke2022} EBL model. The curvature is characterized by a spectral index $\Gamma$=1.80$\pm$0.08 and a curvature coefficient of $\beta$=0.27$\pm$0.08 for the LPxEBL \textit{finke2022} model, which is also compatible with the results of the LPxEBL with \textit{dominguez-upper} and \textit{kneiske} EBL models. Although the spectral curvature is somewhat more significant in analysis of the full energy range in comparison to the reduced energy range (Table~\ref{table:all_fit_results}), they are compatible with each other.

In this study, we also investigate the degeneracy between the EBL absorption feature and an intrinsic curved spectrum (Fig.~\ref{fig:contour}). We conclude that, despite not being able to separate both dependencies completely, the spectral curvature is detected even when considering the EBL with normalization as predicted by the models ($\alpha_\mathrm{norm}$=1). Furthermore, we derive the maximum allowed EBL normalization $\alpha_\mathrm{norm}$ within 95\% confidence level of $\alpha_\mathrm{norm}$<8.7, $\alpha_\mathrm{norm}$<5.5, $\alpha_\mathrm{norm}$<2.0, for the \textit{kneiske}, \textit{finke2022}, and \textit{dominguez-upper} EBL models, respectively. In addition to the LPxEBL model, we also fitted the high-state data with a spectrum described by an PL with exponential cut-off (Appendix~\ref{app:EC}), which leads to a slightly worse description of the data.

In summary, we measure for the first time a curvature in the VHE gamma-ray spectrum of the high state of M\,87. We rule out external absorption by EBL photons as its sole sources and we deem multiple particle populations in the high state as an unlikely explanation for the detected curvature. Furthermore, our estimates show that absorption by star light at 1.6$\,$$\mu m$ is minimal and unlikely to explain the reported curvature. Finally, intrinsic curved radiation spectrum, i.e. due to a cut off in the particle spectrum at the particle accelerator or due to the onset of the Klein-Nishina effect, also stand as plausible explanations for the observed curvature.


\begin{acknowledgements}
The support of the Namibian authorities and of the University of Namibia in facilitating the construction and operation of H.E.S.S. is gratefully acknowledged, as is the support by the German Ministry for Education and Research (BMBF), the Max Planck Society, the German Research Foundation (DFG), the Helmholtz Association, the Alexander von Humboldt Foundation, the French Ministry of Higher Education, Research and Innovation, the Centre National de la Recherche Scientifique (CNRS/IN2P3 and CNRS/INSU), the Commissariat à l’énergie atomique et aux énergies alternatives (CEA), the U.K. Science and Technology Facilities Council (STFC), the Irish Research Council (IRC) and the Science Foundation Ireland (SFI), the Knut and Alice Wallenberg Foundation, the Polish Ministry of Education and Science, agreement no. 2021/WK/06, the South African Department of Science and Technology and National Research Foundation, the University of Namibia, the National Commission on Research, Science \& Technology of Namibia (NCRST), the Austrian Federal Ministry of Education, Science and Research and the Austrian Science Fund (FWF), the Australian Research Council (ARC), the Japan Society for the Promotion of Science, the University of Amsterdam and the Science Committee of Armenia grant 21AG-1C085. We appreciate the excellent work of the technical support staff in Berlin, Zeuthen, Heidelberg, Palaiseau, Paris, Saclay, Tübingen and in Namibia in the construction and operation of the equipment. This work benefited from services provided by the H.E.S.S. Virtual Organisation, supported by the national resource providers of the EGI Federation.

\end{acknowledgements}
\bibliographystyle{aa}
\bibliography{references.bib}

\begin{appendix}

\section{EBL models and derived ULs}

\begin{figure}[hb!]
    \centering
    \includegraphics[width=\textwidth/2]{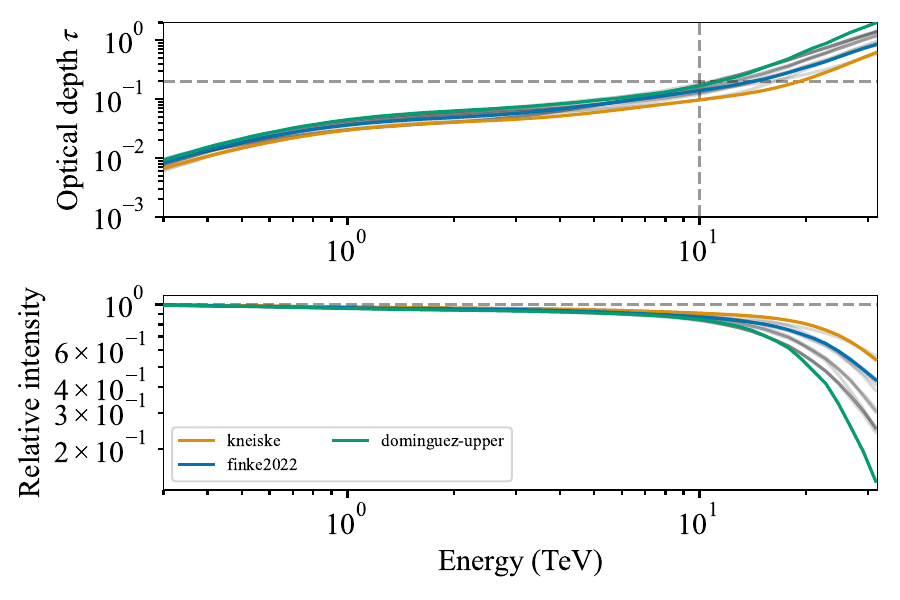}
    \caption{The optical depth (top) and the relative intensity of the gamma-ray flux after absorption (bottom) at z$\approx$0.0042 of the three EBL models employed in this study are presented based on the information provided in the table from \citet{manuel}. The two dashed lines in the upper panel indicate the energy and attenuation levels at which we deem the influence of the EBL spectrum on the overall spectrum as non-negligible. Additionally, for comparison, the remaining available EBL models are depicted in faded black.  In the bottom panel, a reference dashed line is provided to guide the eyes in the case of no absorption.}
    \label{app:fig:tau}
\end{figure}

In this section, we delve into the optical depths of the three EBL models employed in this study and present the derived upper limits (ULs). The chosen models were selected to encompass a wide range of potential attenuation scenarios and are representative, as illustrated in Fig.~\ref{app:fig:tau}, within the gamma-ray energy range considered in our investigation.

\begin{figure}
    \centering
    \includegraphics[width=\textwidth/2]{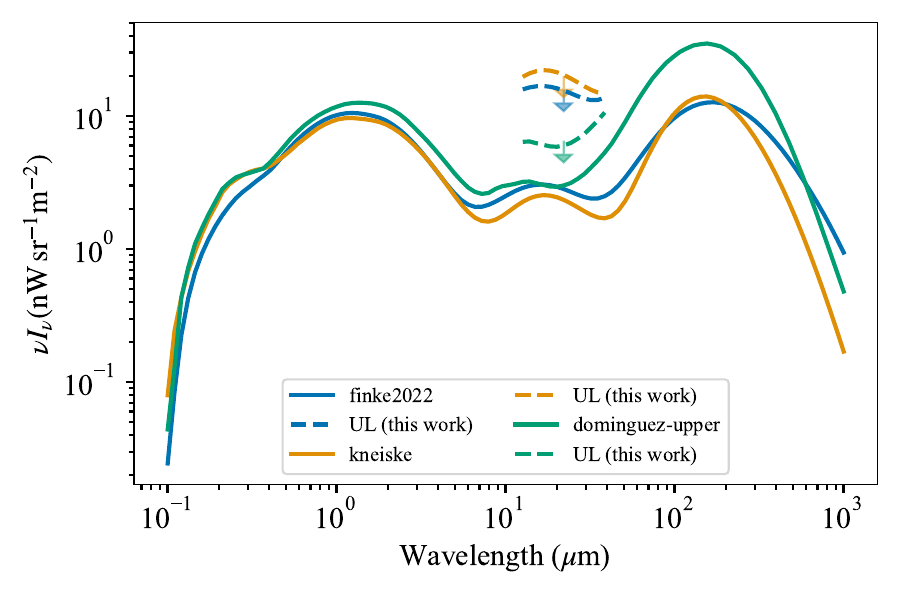}
    \caption{The EBL photon distribution based on the three models employed in this study - \textit{kneiske}, \textit{finke2022}, and \textit{dominguez-upper} models - is depicted in orange, blue, and green, respectively. The 95\% confidence level upper limits (ULs) derived from fitting the high state, as defined in the main text, to LPxEBL models are represented by dashed lines.}
    \label{app:fig:EBL}
\end{figure}

Based on the results obtained from fitting the LPxEBL models to the high-state data of M,87 and exploring the interplay between the curvature parameter and the EBL normalization, we have derived, in the main text, 95\% confidence level upper limits (ULs) on $\alpha_\mathrm{norm}$. Concluding this study, in Fig.~\ref{app:fig:EBL}, we present the spectral distribution of the EBL photons along with the corresponding ULs. The ULs on the EBL distribution are showcased within the gamma-ray energy range of 10$\,$TeV to 32$\,$TeV (EBL photon energy range between 12.4 and 40$\,\mu$m), where our work exhibits the highest sensitivity. It is noteworthy that our strongest UL pertains to the brightest EBL model (\textit{dominguez-upper}). Despite not very constraining, ULs in these regions of the EBL spectrum are relatively uncommon \citep{Biteau2015}. 

\section{Systematic analysis of the high-state selection}
\label{app:spectra}
\begin{figure}
    \centering
    \includegraphics[width=\textwidth/2]{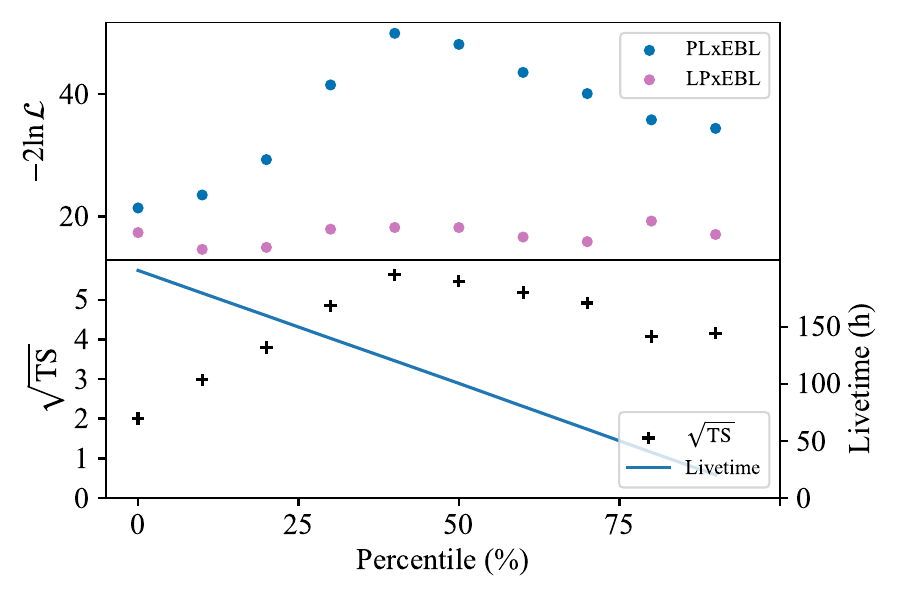}
    \caption{The TS distribution for the PLxEBL (blue) and LPxEBL (purple) models, incorporating the \textit{finke2022} EBL model, is analyzed in relation to datasets defined by the flux distribution above 1$\,$TeV. Each dataset encompasses observations with flux surpassing a specified percentile. The lower panel presents the distribution of $\sqrt{TS}$ denoted by black crosses, while the solid line illustrates the dataset's livetime distribution.}
    \label{fig:app:TS}
\end{figure}

The choice of the high-state dataset significantly influences the outcomes of the final spectral fitting results. To explore this influence, we established nine distinct datasets. Each dataset comprises observations with a flux above 1$\,$TeV specifically exceeding a predefined percentile ranging from 0\% to 90\%, with intervals of 10\%. Subsequently, we applied fits to these datasets using PLxEBL and LPxEBL models, employing the \textit{finke2022} EBL model. The results, depicting the likelihood distribution and $\sqrt{{TS}}$ of the curved model in comparison to the PL model, are presented in Fig.~\ref{fig:app:TS}. Notably, we observe the detection of curvature at a significance level exceeding 4$\,\sigma$, commencing with datasets featuring flux above the 30\% percentile and extending to those above the 90\% percentile. Moreover, the curvature is evident at a 5$\,\sigma$ level in datasets ranging from flux above the 30\% percentile to those above the 60\% percentile. Consequently, we consider the measurement of curvature in the highest flux observation runs to be robust. Variations in $\sqrt{{TS}}$ stem from both event statistics, which systematically decrease with higher cuts, and the presence of low-state runs in the dataset. The latter is particularly relevant for datasets up to the 40\% percentile cut.

\begin{table*}[ht!]
    \centering
    \begin{tabular}{c c c c c c c}
        \hline
        \hline
        Model &
        -2ln\(\mathcal{L}\) &
        $\mathrm{\phi_0}$ &
        \textbf{$\Gamma$} &
        $E_\mathrm{c}$ & 
        Ref. energy & 
        Energy range \\
        \hline
        PLxECxEBL \textit{finke2022} & 18.0 & 11$\pm$3 & 1.02$\pm$0.26 & 3$\pm$1 & 1.96 & 0.3 -- 10 \\
        PLxECxEBL \textit{finke2022} & 20.8 & 5.9$\pm$0.9 & 1.45$\pm$0.16 & 8$\pm$3 & 2.15 & 0.3 -- 32 \\
        \hline
            
    \end{tabular}
    \caption{Best fit results for the spectral VHE gamma-ray distribution of the high state of M\,87. In addition to the models presented in Table~\ref{table:all_fit_results}, we include the best-fit PLxECxEBL \textit{finke2022} for the reduced energy range and for full energy range. The energy range and the critical energy are given in TeV and $\mathrm{\phi_0}$ is given in units of 10$^{-13}$cm$^{-2}$ s$^{-1}$TeV$^{-1}$. The decorrelation energy of the PL spectral model was used as the reference energy, given in TeV in the table.}
    \label{app:table:fit_results}
\end{table*}

\begin{figure}[ht!]
    \centering
    \includegraphics[width=\textwidth/2]{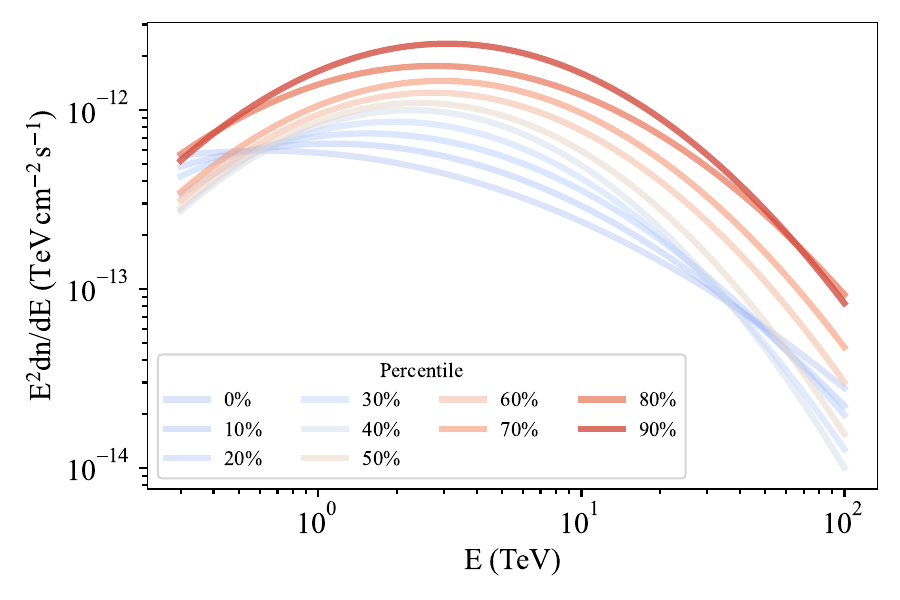}
    \caption{The best spectral fits for the LPxEBL models are illustrated for the various datasets, revealing the emergence of a curved spectrum in the high state.}
    \label{fig:app:spectra}
\end{figure}

Fig.~\ref{fig:app:spectra} displays the best-fit LPxEBL model for diverse datasets categorized by their percentile cuts. The spectrum derived from the entire dataset (0\% percentile) suggests a tendency for curvature, which becomes more pronounced in datasets focusing solely on observations with the highest fluxes. These outcomes reinforce the previously reported detection of curvature in the high-state VHE gamma-ray emissions of M\,87 in the main text.

To conclude the systematic checks, we conducted an analysis of the high state as defined in \citet{Victor}, utilizing Bayesian block analysis. It is worth noting that this definition may not be ideal for studying high states, given that flares are smoothed out in the presence of nearby low emission states. However, despite this limitation, the spectral fit based on the high state definition exhibits a preference of 3.6$\,$$\sigma$ for an LPxEBL \textit{finke2022} model compared to the PLxEBL \textit{finke2022} model. Importantly, this result is within 1$\,$$\sigma$ consistency with the findings presented in the main text.

\section{The spectrum according to an exponential cut-off function}
\label{app:EC}

The emergence of both the Klein-Nishina regime and the spectral region close to the maximum particle energy achievable by the accelerator might give rise to an exponential cut-off gamma-ray distribution in a synchrotron-self Compton scenario, potentially resembling curvature. To explore this possibility, we define a PL with an exponential cut-off model including the EBL absorption (PLxECxEBL \textit{finke2022}):

\begin{equation}
\label{eq:EC}
    \phi(E) = \phi_0 \left( \frac{E}{E_0} \right)^{- \Gamma} \exp \left [- \left( \frac{E}{E_c} \right) ^{\beta_{c}}  - \alpha_{\text{norm}}\tau(E,z) \right ],
\end{equation}

\noindent where $E_\mathrm{c}$ is the critical energy and $\beta_\mathrm{c}$ determines the steepness of the cut-off.

We fitted the high state data as defined in the main text to Eq.~\ref{eq:EC} with fixed $\alpha_{\text{norm}}=1$ and fixed $\beta_\mathrm{c}=1$ both for the reduced (0.3$\,$-$\,$10$\,$TeV) and for the complete (0.3$\,$-$\,$32$\,$TeV) energy range. The results are presented in  Table~\ref{app:table:fit_results}. In the reduced energy range, the fit statistic of the best-fit PLxECxEBL model is indistinguishable from the best-fit LPxEBL model (Table~\ref{table:all_fit_results}). However, the spectral index of $\Gamma$=$1.0\pm0.3$ appears exceptionally hard for an intrinsic PL distribution and likely nonphysical. In the full energy range, despite the more realistic best-fit spectral index, the fit statistic of the PLxECxEBL model is marginally poorer than that of the LPxEBL \textit{finke2022} model in the same energy range, implying that a cut-off with $\beta_\mathrm{c}=1$ is not the preferred explanation for the detected curvature in the VHE gamma-ray high state of M\,87. We refrain from fitting more complex models, i.e. a PLxECxEBL model with free $\beta_\mathrm{c}$, as such models inevitably lead to degeneracy and inconclusive results for the current dataset with limited event statistics.

\end{appendix}

\end{document}